\documentclass{iopart}
\usepackage{iopams}
\usepackage{graphicx}
\begin{document}

\vspace*{-3.cm}
\begin{flushright}
TUM-HEP-477/02\\
\end{flushright}

\title{The effects of matter density uncertainties on neutrino oscillations in
the Earth}

\author{Bj{\"o}rn Jacobsson\dag, Tommy Ohlsson\dag, H{\aa}kan Snellman\dag,
and Walter Winter\ddag\footnote[3]{Speaker}}

\address{\dag\ Division of
Mathematical Physics, Department of Physics, Royal Institute of
Technology - Stockholm Center for Physics, Astronomy, and
Biotechnology, \newline 106~91 Stockholm, Sweden}

\address{\ddag\ Institut f{\"u}r Theoretische
Physik, Physik-Department, Technische Universit{\"a}t M{\"u}nchen,
James-Franck-Stra{\ss}e, 85748 Garching bei M{\"u}nchen,
Germany}

\begin{abstract}
We compare three different methods to evaluate uncertainties in the
Earth's matter density profile, which are relevant to long baseline
experiments, such as neutrino factories.
\end{abstract}



%\maketitle

%\section{Introduction}

It is generally believed that neutrino oscillations are influenced by
the presence of matter~\cite{mikh85,mikh86,wolf78}. Especially, for very long
baseline experiments, such as often proposed for neutrino factories, matter
effects can be quite substantial. The knowledge on the matter density profile
along a certain baseline through the Earth's mantle is limited to about $5 \%$
precision (for a summary, see, e.g., Ref.~\cite{Pana}). Thus, matter density
uncertainties can affect the measurement of the neutrino oscillation parameters,
such as the CP phase $\delta_{CP}$~\cite{Shan:2001br}. There have been several
approaches to model the matter density uncertainties and check their influence
on neutrino
oscillations~\cite{Shan:2001br,Jacobsson:2001zk,Shan:2002px,Huber:2002mx}. In
this talk, we especially focus on three different techniques and show their
advantages and disadvantages where applicable. Based on these techniques, we
will finally comment on the relevance of matter density uncertainties compared
to other problems. %
\begin{figure}[ht!]
\caption{\label{ourmodels} The different models for matter density
uncertainties used in this presentation for a baseline length of $7400 \,
\mathrm{km}$: a single perturbation (left), random fluctuations (middle),
and the mean density as measured quantity (right).} \begin{center}
\includegraphics[width=13cm]{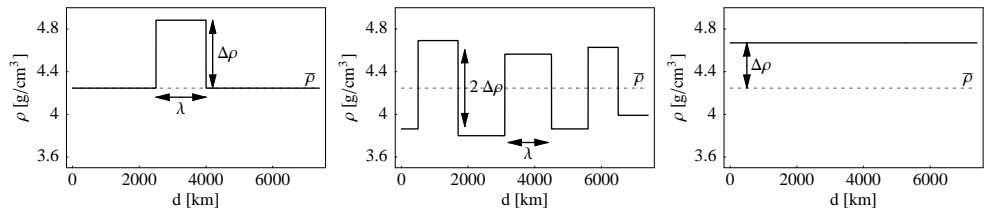}
\end{center}
\end{figure}
Figure~\ref{ourmodels} shows the models for matter density uncertainties used in
this presentation.
The left-hand plot illustrates a single perturbation of the average matter
density, such as it could come from a mine, mountain, or lake. The middle plot
shows random fluctuations, which can be used to model realistic fluctuations
known from geophysics. The right-hand plot illustrates the technique of
measuring the average matter density, i.e., the density difference $\Delta
\rho$ between the average matter density from the PREM profile~\cite{dzie81}
and the actual average density is measured as an additional quantity by the
neutrino oscillation experiment.

%\section{A single perturbation}

As indicated in Fig.~\ref{ourmodels} (left), a single
perturbation could be modeled as a bump with a length scale $\lambda$ and an
amplitude $\Delta \rho$. A general result is that the relative error in the
appearance probability of a neutrino factory is smaller than about $1 \%$ for
applications such as coal mines and lakes with a length scale $\lambda \sim 10
\, \mathrm{km}$ and a density contrast $\Delta \rho/\rho \lesssim 100 \%
$~\cite{Jacobsson:2001zk} (for the oscillation parameters, we choose,
if not otherwise stated, the LMA solution, $L=7400 \, \mathrm{km}$, and $E=
30 \, \mathrm{GeV}$). Analytically, one can show with perturbation theory that
the change in the probabilities is proportional to the product $\lambda \,
\Delta \rho$~\cite{Jacobsson:2001zk}.

%\section{Random density fluctuations}

In geophysics, uncertainties in the matter density profile up to 5 \% have been
documented. Unfortunately, there is no general agreement among the geophysics
results. However, one can easily observe common characteristics of measurements
in certain depths~(see, e.g., Ref.~\cite{Pana}), such as the length scales and
amplitudes of the fluctuations. With the random density fluctuations method
we model the matter density uncertainties based on their characteristics and
average over a large number of random matter density profiles in order to
estimate the average effects~\cite{Jacobsson:2001zk}. As indicated in
Fig.~\ref{ourmodels} (middle plot), we introduce two different parameters, which
are again the length scale $\lambda$ and the amplitude $\Delta \rho$. These we
allow to vary around their average values with Gaussian distributions with
standard deviations $\sigma_\lambda$ and $\sigma_{\Delta \rho}$, respectively.
From geophysics maps, we can estimate the parameter values: $\lambda \sim 2000
\, \mathrm{km}$, $\sigma_\lambda \sim 1500 \, \mathrm{km}$, $\Delta \rho \sim 3
\% \, \bar{\rho}$, and $\sigma_{\Delta \rho} \sim 1 \% \, \bar{\rho}$. As the
most interesting result, we find that the errors in the appearance
probabilities qualitatively behave as the ones for a single perturbation,
i.e., they are essentially proportional to the product of the length scale and
the amplitude. However, because of averaging effects, they are suppressed by a
factor of about two to three. Note that they are not necessarily completely
averaging out because of interference effects: It is easy to show that the
quantum mechanical evolution operators in different density layers do
not commute. Analytically, one can again show with perturbation theory that fast
fluctuations on length scales much shorter than the oscillation length in
matter average out for limited amplitudes~\cite{Ohlsson:2001ck}.

%\section{Mean density as measured quantity}

A different approach is to assume the mean density to be known 
with about $\pm 5 \%$ precision and to measure it, within these limits, together
with the other neutrino oscillation parameters by the experiment (cf.,
Fig.~\ref{ourmodels}, right plot). This method can be applied to a full
statistical neutrino factory analysis, such as it is done in
Ref.~\cite{Huber:2002mx} for an initial and advanced stage neutrino factory,
respectively, at a baseline length of $3000 \, \mathrm{km}$ ($0.75 \,
\mathrm{MW}$ and $4 \, \mathrm{MW}$ target power, $5 \, \mathrm{yr}$ and $8 \,
\mathrm{yr}$ running time, $10 \, \mathrm{kt}$ and $50 \, \mathrm{kt}$ detector
mass, respectively). However, it should be noted that this method in many cases
only allows a quite conservative estimate, since it does not take into account
averaging effects among more than one density layer. As the main result of this
method, the effects of the matter density uncertainties are most important for
the measurement of the CP phase at the advanced stage at a neutrino factory.
For the measurement of $\sin^2 2 \theta_{13}$, many other experimental issues,
such as the efficiencies at low energies, are of comparable magnitude.

%\section{Summary and conclusions}

Let us summarize the individual approaches. The single perturbation
method can be evaluated analytically on the level of oscillation probabilities.
The relative errors are proportional to the product $\lambda \, \Delta
\rho$ and no averaging effects enter this result. The errors are small for
lakes, mountains, mines, etc., and the method has only limited applications.
The random fluctuations approach can be used numerically
on the level of neutrino oscillation probabilities. The errors on the
neutrino oscillation probabilities are qualitatively proportional to the product
of $\lambda$ and $\Delta \rho$. Since there are averaging effects, they are,
however, suppressed by a factor of about two or three compared to the single
perturbation. For large enough structures, such as for tectonic plates or
realistic fluctuations in the Earth's mantle, the impact on the results can be
much larger than the one of a single perturbation. One problem is the high
computational effort, which means that it is hard to apply this method in a
complete statistical analysis. However, in the limit of large enough $\lambda
\sim L$, the next method provides a good approximation:
The mean density as measured quantity approach can be used in a complete
statistical analysis, which needs to be performed numerically. The errors on the
neutrino oscillation probabilities can then be directly translated into the
errors on the quantities to be measured by the experiment. Especially, a CP
phase measurement can be substantially affected by matter density uncertainties.
Averaging and interference effects of matter density fluctuations are, however,
not directly modeled by this method. In this context,
assuming a $5 \%$ error on the average matter density should be a rather
conservative estimate, since averaging effects are completely neglected.

We have seen that there can be quite substantial effects of matter density
uncertainties in a complete statistical analysis. However, we have demonstrated
that these are partially reduced by averaging in a more realistic random
fluctuations model. Thus, we may expect that the effects in a complete analysis
are smaller than for the conservative estimate of the measured average
matter density. In addition, the information on the matter density profile
along a specific baseline from geophysics should be much better than assumed
here, since at least a part of the terrain in lower depths should be well-known
and the existing information could be combined for higher depths. Furthermore,
matter density uncertainties are only relevant for a very advanced
neutrino factory experiment to be built at a time when geophysics research has
also been advancing. We conclude that matter density uncertainties will
probably not be the bottleneck of the statistical analysis of a planned
experiment, though somewhat more effort should be spent on improving the results
from geophysics.

This work was supported by the 4th NuFact '02 Workshop and its sponsors,
the Magnus Bergvall Foundation (Magn. Bergvalls Stiftelse), the
``Studienstiftung des deutschen Volkes'' (German National Merit
Foundation), and the ``Sonderforschungsbereich 375 f{\"u}r
Astro-Teilchenphysik der Deutschen Forschungsgemeinschaft''.


\section*{References}

\begin{thebibliography}{10}

\bibitem{mikh85}
S. Mikheyev and A. Smirnov,
\newblock Yad. Fiz. 42 (1985) 1441,
\newblock [Sov. J. Nucl. Phys. 42 (1985) 913].

\bibitem{mikh86}
S. Mikheyev and A. Smirnov,
\newblock Nuovo Cimento C 9 (1986) 17.

\bibitem{wolf78}
L. Wolfenstein,
\newblock Phys. Rev. D17 (1978) 2369.

\bibitem{Pana}
S.V. Panasyuk,
\newblock {REM (Reference Earth Model) web page},
\newblock \newline {\tt http://cfauvcs5.harvard.edu/lana/rem/index.htm}.

\bibitem{Shan:2001br}
L.Y. Shan, B.L. Young and X. Zhang,
\newblock hep-ph/0110414.

\bibitem{Jacobsson:2001zk}
B. Jacobsson et~al.,
\newblock Phys. Lett. B532 (2002) 259, hep-ph/0112138.

\bibitem{Shan:2002px}
L.Y. Shan and X.M. Zhang,
\newblock Phys. Rev. D65 (2002) 113011.

\bibitem{Huber:2002mx}
P. Huber, M. Lindner and W. Winter,
\newblock Nucl. Phys. B  (to be published), hep-ph/0204352.

\bibitem{dzie81}
A.M. Dziewonski and D.L. Anderson,
\newblock Phys. Earth Planet. Inter. 25 (1981) 297.

\bibitem{Ohlsson:2001ck}
T. Ohlsson and W. Winter,
\newblock Phys. Lett. B512 (2001) 357, hep-ph/0105293.

\end{thebibliography}
\end{document}